%% file: eprint.tex
\documentclass[12pt,a4paper]{article}
\usepackage{graphicx}
\usepackage{authblk}
\usepackage{setspace}
\usepackage{subfigure}
\renewcommand\thefootnote{\alph{footnote}}
\def\Tsukuba{University of Tsukuba, Ibaraki 305-8571, Japan}
\def\Jaxa{Institute of Space and Astronautical Science, JAXA, Kanagawa 252-5210, Japan}
\def\Okayama{Okayama university, Okayama 700-8530, Japan}
\def\Riken{RIKEN, Saitama 351-0198, Japan}
\def\Fukui{University of Fukui, Fukui 910-8507, Japan}
\def\Kinki{Kinki University, Osaka 577-8502, Japan}
\def\KEK{High Energy Accelerator Research Organization (KEK), Ibaraki 305-0801, Japan}
\def\Fermi{Fermi National Accelerator Laboratory Batavia, IL 60510-5011, USA}
\def\Seoul{Seoul National University, Seoul 151-742, Korea}
\def\AIST{National Institute of Advanced Industrial Science and Technology (AIST), Ibaraki 305-8568, Japan}
\def\support{\footnote{E-mail address: kkiuchi@hep.px.tsukuba.ac.jp}}

\def\Title#1{\begin{center} {\Large #1 } \end{center}}
\def\Author#1{\begin{center}{ \sc #1} \end{center}}
\def\Address#1{\begin{center}{ \it #1} \end{center}}

\newenvironment{Abstract}{\begin{quotation}  }{\end{quotation}}
\newenvironment{Presented}{\begin{quotation} \begin{center} 
             PRESENTED AT\end{center}\bigskip 
      \begin{center}\begin{large}}{\end{large}\end{center} \end{quotation}}
\def\Acknowledgements{\bigskip  \bigskip \begin{center} \begin{large}
             \bf ACKNOWLEDGEMENTS \end{large}\end{center}}
\input econfmacros.tex
\begin{document}
\begin{titlepage}
%%\pubblock

\vfill
\Title{Development FD-SOI MOSFET amplifiers for integrated read-out circuit of superconducting-tunnel-junction single-photon-detectors}
\vfill
%%\Author{ Kenji Kiuchi\support}
\Author{Kenji Kiuchi$^{1}$\support, Shinhong Kim$^{1}$, Yuji Takeuchi$^{1}$, Kenichi Takemasa$^{1}$, Kazuki Nagata$^{1}$, Kota Kasahara$^{1}$, Koya Moriuchi$^{1}$, Ren Senzaki$^{1}$, Shunsuke Yagi$^{1}$, 
Hirokazu Ikeda$^{2}$, Shuji Matsuura$^{2}$, Takehiko Wada$^{2}$, Hirokazu Ishino$^{3}$, Atsuko Kibayashi$^{3}$, Hiromi Sato$^{4}$, Satoru Mima$^{4}$, Takuo Yoshida$^{5}$, Ryuta Hirose$^{5}$, 
Yukihiro Kato$^{6}$, Masasi Hazumi$^{7}$, Yasuo Arai$^{7}$, Ikuo Kurachi$^{7}$, Erik Ramgerg$^{8}$, Mark Kozlovsky$^{8}$, Paul Rubinov$^{8}$, Dmitri Sergatskov$^{8}$, Soo-Bong Kim$^{9}$,
Shigetomo Shiki$^{10}$, Masahiro Ukibe$^{10}$, Go Fujii$^{10}$ and Masataka Okubo$^{10}$}
\Address{$^{1}$\Tsukuba}
\Address{$^{2}$\Jaxa}
\Address{$^{3}$\Okayama}
\Address{$^{4}$\Riken}
\Address{$^{5}$\Fukui}
\Address{$^{6}$\Kinki}
\Address{$^{7}$\KEK}
\Address{$^{8}$\Fermi}
\Address{$^{9}$\Seoul}
\Address{$^{10}$\AIST}
\vfill
\begin{Abstract}
    We proposed a new high resolution single photon infrared spectrometer for search for radiative decay of cosmic neutrino background(C$\nu$B).
    The superconducting-tunnel-junctions(STJs) are used as a single photon counting device.
    Each STJ consists of Nb/Al/$\mathrm{Al}_x\mathrm{O}_y$/Al/Nb layers and 
    their thicknesses are optimized for the operation temperature at 370~mK cooled by a ${}^{3}$He sorption refrigerator.
    Our STJs achieved the leak current 250~pA and the measured data implies that a smaller area STJ fulfills our requirement.
    FD-SOI MOSFETs are employed to amplify the STJ signal current in order to increase signal-to-noise ratio(S/N).
    FD-SOI MOSFETs can be operated at cryogenic temperature of 370~mK, which reduces the noise of the signal amplification system.
    FD-SOI MOSFET characteristics are measured at cryogenic temperature.
    The Id-Vgs curve shows a sharper turn on with a higher threshold voltage and the Id-Vds curve shows a non linear shape in linear region at cryogenic temperature.
    Taking into account these effects, FD-SOI MOSFETs are available for read-out circuit of STJ detectors.
    The bias voltage for STJ detectors are 0.4~mV and it must be well stabilized to deliver high performance.
    We proposed an FD-SOI MOSFET based charge integrated amplifier design as a read-out circuit of STJ detectors.
    The requirements for an operational amplifier used in the amplifier is estimated using SPICE simulation.
    The op-amp required to have a fast response(GBW$\geq$100~MHz) and it must have low power dissipation as compared to the cooling power of refrigerator.
\end{Abstract}
\vfill
\begin{Presented}
International Workshop on SOI Pixel Detector (SOIPIX2015), Tohoku University, Sendai, Japan, 3-6, June, 2015.
\end{Presented}
\vfill
\end{titlepage}
\def\thefootnote{\fnsymbol{footnote}}
\setcounter{footnote}{0}

\section{Introduction}
S.Kim. et al. proposed an experiment which is to search for cosmic neutrino background decay~\cite{NeutrinoDecay}.
The final goal of this experiment is to measure the mass of neutrino.
Various neutrino oscillation experiments measured the neutrino mass squared differences and 
measured neutrino mass squared difference between $m_{2}$ and $m_{3}$ is measured to be an equation~\ref{eq:nms}~\cite{Neu_OSC}.
\begin{eqnarray}
  \Delta{m_{32}^{2}}=m^{2}_{3}-m^{2}_{2}=(2.44\pm0.06)\times10^{-3}~\mathrm{eV}^{2}
  \label{eq:nms}
\end{eqnarray}
These results indicate that the neutrinos have non zero mass and heavier neutrino($\nu_{3}$) can decay into lighter neutrino($\nu_{2}$) with photon.
The energy of photon from neutrino radiative decay is written as equation~\ref{eq:ph_E}.
\begin{eqnarray}
  E_{\gamma}=\frac{m^{2}_{3}-m^{2}_{2}}{2m_{3}}
  \label{eq:ph_E}
\end{eqnarray}
Once the energy of this photon is measured, the mass of neutrino($m_{3}$) is easily calculated since numerator is already measured as equation~\ref{eq:nms}.
It seems difficult to measure the energy of photon taking in to account extreamly long lifetime of neutrino.
%%%%%%%%%%%%%%%%%%%%%%%%%%%%%%%%%%%%%%%%%%%%%%%%%%%%%%%%%%%%%%%%%%%%%%%%%
\begin{figure}[h]
\begin{center}
\includegraphics[height=2.5cm, keepaspectratio,clip]{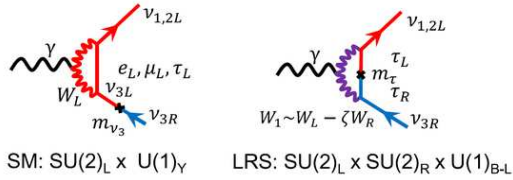}
\caption{$\nu_3 \to \nu_2 + \gamma$ Feynman Diagram in SM(left) and in LRSM(right).}
\label{fig:NeutrinoDecay}
\end{center}
\end{figure}
%%%%%%%%%%%%%%%%%%%%%%%%%%%%%%%%%%%%%%%%%%%%%%%%%%%%%%%%%%%%%%%%%%%%%%%%%
Figure~\ref{fig:NeutrinoDecay} left shows a Feynman diagram of $\nu_3\ \rightarrow\ \nu_2$ neutrino radiative decay in
the Standard Model.
Assuming $m_{3}=50~\mathrm{meV}/c^2$, the lifetime of $\nu_{3}$ is predicted to be in the order of $10^{43}$~years with this process.
Although, in the Left Right Symmetric model(LRSM), figure~\ref{fig:NeutrinoDecay} right process also contributes to the neutrino radiative decay.
This additional process drastically enhances the decay rate and it is predicted to be the order of $10^{17}$~years with the values from recent experimental limits~\cite{NeutrinoDecay}.
The cosmic neutrino background(C$\nu$B) which fills the universe with a particle density $\rho$ of 110~$\mathrm{cm}^{-3}$~per~generation is an ideal neutrino source,
since it provides uniform temperature(1.9~K) enormous number of neutrinos.
The spectrum of C$\nu$B decay is well known.
The energy of the photons from the C$\nu$B decay were monochromatic photons when they were emitted 
and the red shift produces a tail on a long wavelength side.
Main background is zodiacal emission(ZE) which is produced by thermal emission from space dust in the solar system.
The magnitude of zodiacal emission is $\mathcal{O}(10^2)$ larger than the C$\nu$B decay spectrum.
The C$\nu$B decay is measureable thanks to the well known spectrum.
Actually, S.Kim. et al. measured the lifetime of neutrinos with this method and set the current experimental limit on the lifetime.
It is in the order of $10^{12}$~years~\cite{NeutrinoDecay2}. We plan a rocket experiment for improving this limit to $10^{14}$~years as a next step.
We proposed a new high-resolution infrared spectrometer for this experiment.
Our detector system consists of three components similar to the general spectrometer.
Incident infrared photons(40~um - 80~um) are spacially distributed depending on their energy by the diffractive grating.
Superconducting tunnel junctions(STJs) are employed to detect the diffracted photons, 
since it can detect the single infrared photons.
Single photon counting method suppress the dark noise contamination on signal to less than or comparable to ZE on sensitivity degradation.
The fully-depleted silicon-on-insulator(FD-SOI) MOSFET based cold amplifiers are used to amplify the signal current from the STJs.
Following sections describe the status of STJ and cold amplifier development.
\section{Superconducting Tunnel Junction}
%%%%%%%%%%%%%%%%%%%%%%%%%%%%%%%%%%%%%%%%%%%%%%%%%%%%%%%%%%%%%%%%%%%%%%%%%
\begin{figure}[h]
\begin{center}
\includegraphics[height=5cm, keepaspectratio,clip]{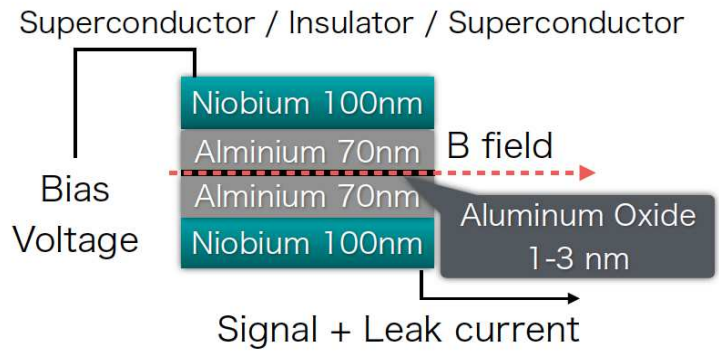}
\caption{Cross-section of our STJ.}
\label{fig:STJ_schematic}
\end{center}
\end{figure}
\begin{figure}[h]
\begin{center}
\includegraphics[height=5cm, keepaspectratio,clip]{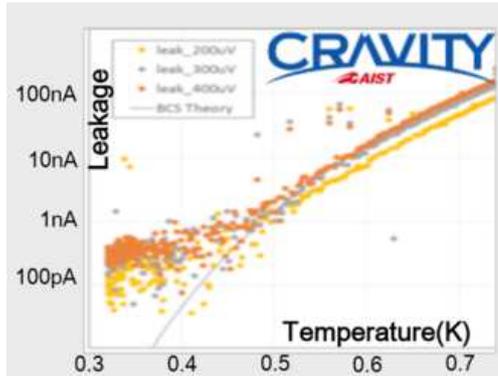}
\caption{Leak currents of 50~$\mu$m$\times$50~$\mu$m STJ.}
\label{fig:STJleak}
\end{center}
\end{figure}

%%%%%%%%%%%%%%%%%%%%%%%%%%%%%%%%%%%%%%%%%%%%%%%%%%%%%%%%%%%%%%%%%%%%%%%%%
Figure~\ref{fig:STJ_schematic} is a cross-section of our STJ.
The STJs were fabricated in the clean room for analog-digital superconductivity (CRAVITY) in AIST.
Each of our STJ consists of Nb/Al/$\mathrm{Al}_x\mathrm{O}_x$/Al/Nb layers.
The electron energy level in the superconductor has a tiny band gap.
An incident particle deposits the energy into the superconductor, exited electrons are produced in the superconductor and 
the number of exited electrons are propotional to the incident energy($N_{e}=E/1.7\Delta$, where $E$ is a deposited energy and $\Delta$ is a band gap energy).
The band gap energy of Al and Nb are 0.175~meV and 1.55~meV, respectively.
There is a proximitiy effect in the superconducting bi-layer, Al/Nb bi-layer has a band gap between pure Al and pure Nb depending on their thicknesses.
Thicknesses of Al and Nb layers are optimized for the operation temperature of 370~mK cooled by the ${}^{3}$He sorption refrigerator.
The $1.7\Delta$ of our Nb/Al STJ is approximately 0.5~meV, this value is 1/2000 of silicon band gap.
This tiny band-gap allows us to detect infrared single photon.
There is an additional current of cooper pairs(supercurrent).
The magnetic field applied parallel to the oxidation layer blocks the supercurrent.
Figure~\ref{fig:STJleak} shows a leak currents of 50~$\mu$m$\times$50~$\mu$m STJ. 
It is measured to be less than 250~pA at the bias voltage of 400~$\mu$V~\cite{STJ}.
This level of leak current already satisfies our requirement.
%%%%%%%%%%%%%%%%%%%%%%%%%%%%%%%%%%%%%%%%%%%%%%%%%%%%%%%%%%%%%%%%%%%%%%%%%
%%%%%%%%%%%%%%%%%%%%%%%%%%%%%%%%%%%%%%%%%%%%%%%%%%%%%%%%%%%%%%%%%%%%%%%%%
\section{FD-SOI based cold amplifier}
\begin{figure}[htbp]
 \begin{minipage}{0.5\hsize}
  \begin{center}
   \includegraphics[width=67mm]{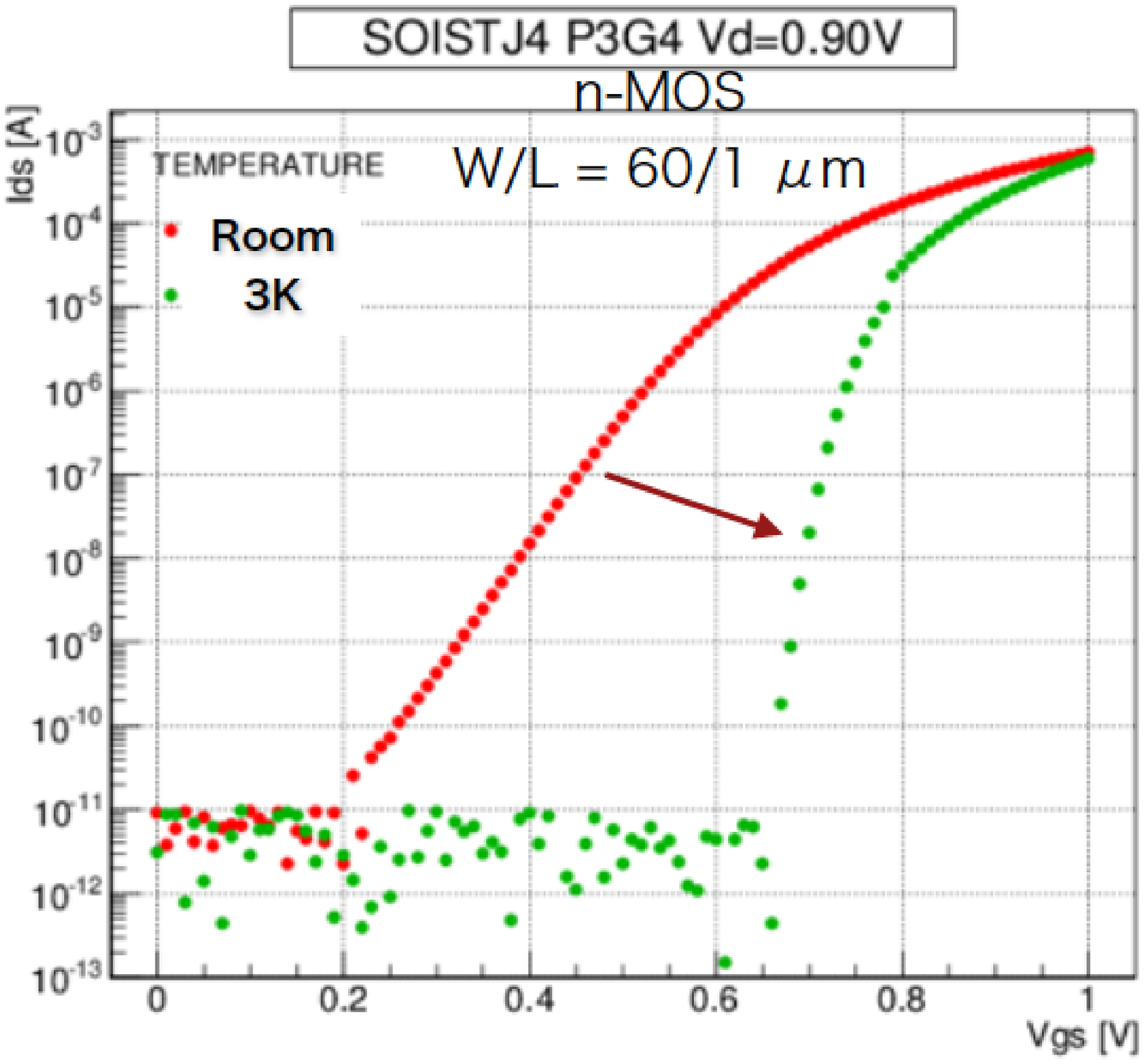}
  \end{center}
  \caption{Id-Vg curve of NMOS \newline source-tie type, W/L = 60/1 $\mu$m}
  \label{fig:IdVg}
 \end{minipage}
 \begin{minipage}{0.5\hsize}
  \begin{center}
   \includegraphics[width=70mm]{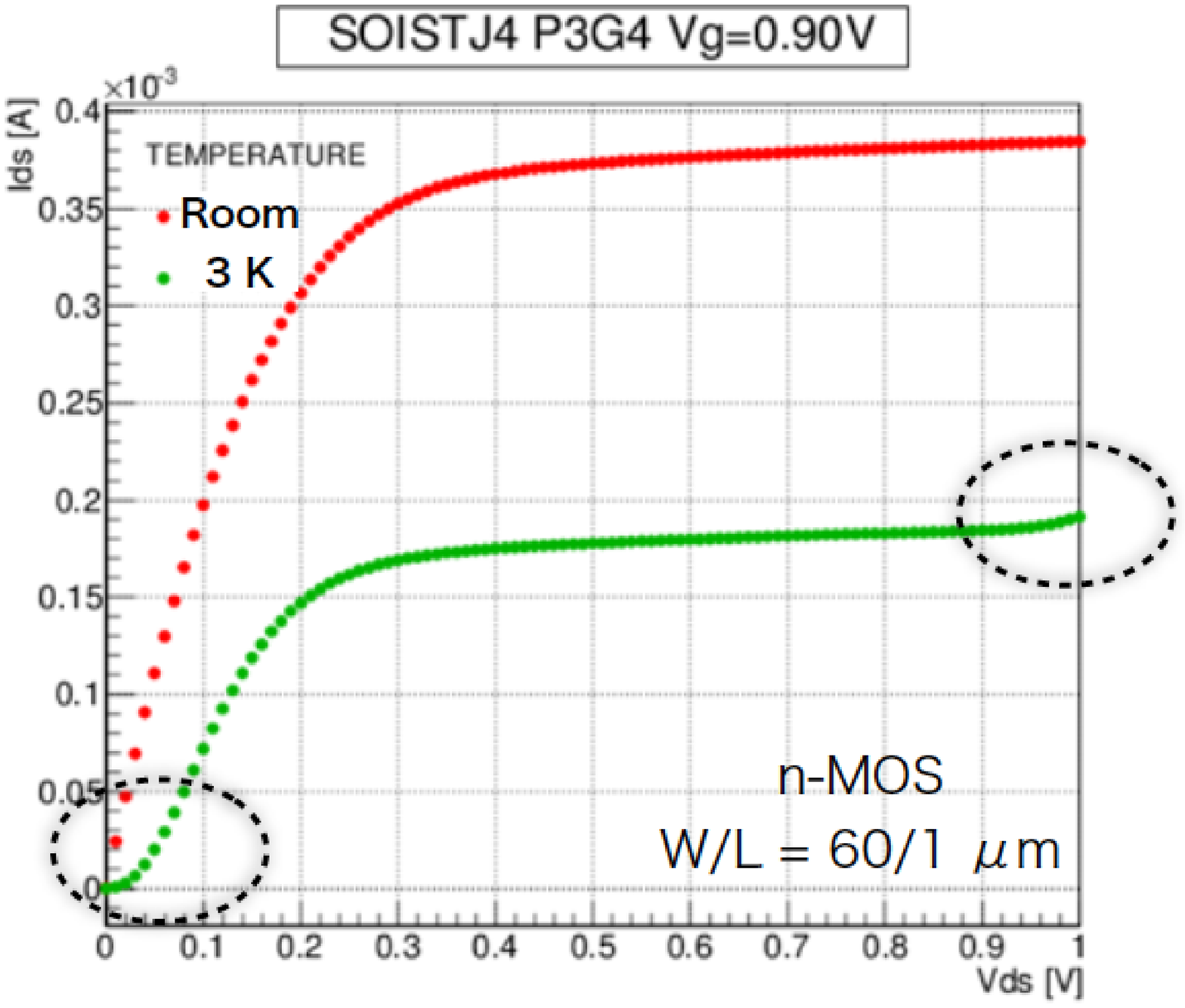}
  \end{center}
  \caption{Id-Vd curve of NMOS \newline source-tie type, W/L = 60/1 $\mu$m}
  \label{fig:IdVd}
 \end{minipage}
\end{figure}

The STJ response to 465~nm photons is measured with conventional charge amplifier.
Typical signal width for 4 photons is $\mathcal{O}(\mu{\mathrm{s}})$. We cannot detect UV single photon at this experiment.
This performance degradation is due to noises(thermal noises, electromagnetic wave absorbtion) and transmission loss in the read out system.
An introduction of a cold amplifier is the best solution to minimize the degradation.
The JFET is well used to the cold amplifier for STJ. However, the operation temperature of JFET is much higher than the operation temperature of STJ.
JAXA/ISAS reported that the FD-SOI MOSFETs shows an excellent performance at liquid helium temperature(4.2~K) and 
we confirmed that they can be operated at the operation temperature less than 100~mK~\cite{soifet}.
The FD-SOI based pre-amplifier amplify the STJ output charge nearby STJ on a temperature stage of 370~mK.
Generally, the carrier density in semiconductor decreases at low temperature due to incomplete ionization of the dopants.
The carrier mobility increases at low temperature due to decreasing of scattering probability.
Hence, the characteristics of FD-SOI MOSFET at cryogenic temperature can be changed compared to the room temperature.
The IV characteristics of FD-SOI MOSFETs at cryogenic temperature are measured to prepare the SPICE models for cryogenic
temperature.
FD-SOI MOSFETs are processed by Lapis semiconductor 0.2~$\mu$m process.
The four wire sensing method is used to avoid the voltage drop due to the readout wires in refrigerator.
Figure~\ref{fig:IdVg} left shows Id-Vg characteristics of NMOS source-tie type with W/L~=~60/1~$\mu$m at Vd=0.9~V.
Threshold is increased and transconductance is increased with decreasing temperature.
These phenomena implies decreasing the carrier density and increasing the carrier mobility as we expected.
Figure~\ref{fig:IdVd} right shows Id-Vd characteristics of the same FET at Vg=0.9~V.
The non-linear shape is observed at low Vds region and the kink effect is found in high Vds.
The cause of the non-linear shape is currently under investigation.
Taking into account these effects, FD-SOI MOSFETs are available for read-out circuit of STJ detectors.
We will design the charge integration amplifier using special SPICE model.
The operational amplifier for this circuit is required to have very fast response(GBW=100~MHz), 
since the width of STJ signal current is a few $\mu$s and STJ has large capacitance(calculated to be $\approx$80fF/$\mu$m$^{2}$).
It must have low power dissipation as compared to the cooling power of refrigerator.
%%%%%%%%%%%%%%%%%%%%%%%%%%%%%%%%%%%%%%%%%%%%%%%%%%%%%%%%%%%%%%%%%%%%%%%%%
%%%%%%%%%%%%%%%%%%%%%%%%%%%%%%%%%%%%%%%%%%%%%%%%%%%%%%%%%%%%%%%%%%%%%%%%%
\section{Conclusion}
We proposed a new high resolution infrared single photon spectrometer for C$\nu$B radiative decay search.
The infrared spectrometer has a dispersive element.
The STJ is employed to detect infrared single photon.
Our STJ achieved the leak current of 250~pA at the bias voltage of 400~$\mu$V.
The FD-SOI MOSFETs are used to design cold amplifier for STJ readout system.
The FD-SOI MOSFETs show excellent performances, non-linear Id-Vd characteristic at low Vd region is under investigation.
We will design the charge integration amplifier using FD-SOI MOSFETs.
%%%%%%%%%%%%%%%%%%%%%%%%%%%%%%%%%%%%%%%%%%%%%%%%%%%%%%%%%%%%%%%%%%%%%%%%%
\newpage
\Acknowledgements
This work was supported by the Ministry of Education, Science, Sports and Culture of Japan (MEXT KAKENHI Grant Number 25105007).
The devices were fabricated in the clean room for analog-digital superconductivity (CRAVITY) in National Institute of Advanced Industrial Science and Technology (AIST).
This work is supported by VLSI Design and Education Center(VDEC), The University of Tokyo with the collaboration with Cadence, Synopsys and Mentor Graphics Corporation.

%%%%%%%%%%%%%%%%%%%%%%%%%%%%%
\end{document}

%% file: econfmacros.tex
\def\beq{\begin{equation}}
\def\eeq#1{\label{#1}\end{equation}}
\def\eeqn{\end{equation}}

\def\beqa{\begin{eqnarray}}
\def\eeqa#1{\label{#1}\end{eqnarray}}
\def\eeqan{\end{eqnarray}}

\let\bar=\overbar

\def\Dslash{\not{\hbox{\kern-4pt $D$}}}
\def\dslash{\not{\hbox{\kern-2pt $\del$}}}

\def\msb{{\bar{\ssstyle M \kern -1pt S}}}